\documentclass[12pt,twoside]{article}   
\usepackage{epsfig,times}  
\usepackage{color}

\begin{document}
\thispagestyle{empty} 


 \renewcommand{\topfraction}{.99}      
 \renewcommand{\bottomfraction}{.99} 
 \renewcommand{\textfraction}{.0}


\newcommand{\nc}{\newcommand}

\nc{\qI}[1]{\section{{#1}}}
\nc{\qA}[1]{\subsection{{#1}}}
\nc{\qun}[1]{\subsubsection{{#1}}}
\nc{\qa}[1]{\paragraph{{#1}}}

\def\qbu{\hfill \par \hskip 6mm $ \bullet $ \hskip 2mm}
\def\qee#1{\hfill \par \hskip 6mm #1 \hskip 2 mm}

\nc{\qfoot}[1]{\footnote{{#1}}}
\def\qL{\hfill \break}
\def\qpar{\vskip 2mm plus 0.2mm minus 0.2mm}
\def\tvi{\vrule height 12pt depth 5pt width 0pt}
\def\qtvi{\vrule height 2pt depth 5pt width 0pt}
\def\qth{\vrule height 15pt depth 0pt width 0pt}
\def\qtb{\vrule height 0pt depth 5pt width 0pt}

\def\qparr{ \vskip 1.0mm plus 0.2mm minus 0.2mm \hangindent=10mm
\hangafter=1}

\def\qdec#1{\par {\leftskip=2cm {#1} \par}}
\def\qbfb#1{{\bf\color{blue}{#1} }}

\def\qdpt{\partial_t}
\def\qdpx{\partial_x}
\def\qddpt{\partial^{2}_{t^2}}
\def\qddpx{\partial^{2}_{x^2}}
\def\qn#1{\eqno \hbox{(#1)}}
\def\qds{\displaystyle}
\def\qw{\widetilde}
\def\qmax{\mathop{\rm Max}}   
\def\qmin{\mathop{\rm Min}}   

\def\qs#1{{\bf \color{blue} \LARGE {#1}}\quad }

\def\qv{\vskip 0.1mm plus 0.05mm minus 0.05mm}
\def\qhu{\hskip 1mm}
\def\qhv{\hskip 3mm}
\def\qvv{\vskip 0.5mm plus 0.2mm minus 0.2mm}
\def\qhw{\hskip 1.5mm}
\def\qleg#1#2#3{\noindent {\bf \small #1\qhw}{\small #2\qhw}{\it \small #3}\qv }


\centerline{\bf \Large 
Transient frailty induced by cell division.}
\centerline{\bf \Large 
Observation, reasons and implications }
\vskip 7mm 
\vskip 10mm

\centerline{\normalsize
Zengru Di$ ^1 $,
Eduardo M. Garcia-Roger$ ^2 $,
Peter Richmond$ ^3 $,}
\qL
\centerline{\normalsize
Bertrand M. Roehner$ ^4 $,
Stephane Tronche$ ^5 $ 
}

\vskip 5mm
\large
                              
\vskip 5mm
\centerline{\it \small Version of 10 December 2021}
\centerline{\it \small Provisional. Comments are welcome.}
\vskip 3mm

{\small Key-words: Escherichia coli, cell division, mortality
rate, frailty} 

\vskip 3mm

{\normalsize
1: School of Systems Science, Beijing Normal University, China.\qL
Email: zdi@bnu.edu.cn\qL
2: Institut Cavanilles de Biodiversitat I Biologia Evolutiva,
University of Val\`encia, Spain.\qL
Email: eduardo.garcia@uv.es\qL
3: School of Physics, Trinity College Dublin, Ireland.\qL
Email: peter\_richmond@ymail.com \qL
4: Institute for Theoretical and High Energy Physics (LPTHE),
Pierre and Marie Curie campus, Sorbonne University,
National Center for Scientific  Research (CNRS),
Paris, France. \qL
Email: roehner@lpthe.jussieu.fr\qL
5: Jean Perrin Laboratory, Pierre and Marie Curie campus,
Sorbonne University, Paris, France.\qL
Email: stephane.tronche@upmc.fr
}
\vfill\eject


\large

{\bf Abstract}
\qpar

We know that stress-factors, e.g. X-rays, have an effect on 
cells that is more lethal in rapid exponential growth
than in stationary phase. It is this effect which 
makes radiotherapy effective in cancer treatment.
This stress effect can be explained in two ways: 
(a) more vulnerability in the growth phase, 
(b) improved protection capacity and repair 
mechanisms in the stationary phase. 
Although the two explanations
do not exclude each other, they are very different in the sense
that (a) is a general mechanism whereas (b) is strain and 
stress-factor dependent. \qL
In this paper we explore major facets of (a).
Firstly, we emphasize that (a) can account for known experimental
stress-factor evidence. Secondly, we observe that (a) rightly
predicts that slow exponential growth (meaning with a doubling time
of several hours) results in a lower death rate than fast 
exponential growth (doubling time of a fraction of one
hour), an effect that cannot be explained in the (b) framework
because both organisms are in the same phase.
Thirdly, we conjecture that the stress-factor effect 
can be extended to situations of {\it chronic stress} due to 
non-optimal environmental conditions. If correct, this conjecture
would imply that even in normal culture conditions the natural
death rate is lower in the stationary phase than in the 
growth phase.  Finally, the paper closes with the description 
of several
open questions and of appropriate test-experiments
meant to address them.

\qI{Introduction}

To facilitate reading it can be observed that this paper was
written in what can be called the ``spirit of physics''.
This means that it focuses on
general causes rather than on the details of specific cases. As a
parallel in physics
one can mention the Navier-Stokes equations which describe
the basic features of fluid flows, whether in air, water or
any other fluid.

\qA{Vital shocks}

It is a natural idea to suspect that when a living organism
undergoes a major transition from a state $ A $ to a state  
$ B $ it faces elevated risks until
the adjustment is completed successfully. Note that the change 
may affect the environment of the organism, or/and its
internal state.
An example which comes to mind is the
process of moulding, e.g. when a cicada
gets rid of its former exoskeleton . Although the new
replacement exoskeleton has taken form under the old
one, it remains soft for several hours (or days in case
of larger species). During this short time 
the cicada is more vulnerable to predators.
\qpar

In a series of previous papers (Richmond et al. 2016a,b)
we have considered
transient situations in human demography.
Several cases were described in which an abrupt
change in living conditions leads to a mortality spike.
This effect was summarized in what we called the
``{\it Transient Shock conjecture}''. It provides
a qualitative model which leads to testable predictions. 
For instance, statistical evidence shows that persons
who become widowed experience within months
a  mortality spike
that is higher than their average long-term
death rate either before or after
becoming widowed. This case may not come as a surprise.
More revealing is the marriage transition.
Marriage
certainly brings about a major change in personal and 
social conditions and according to
our conjecture one would expect a mortality spike in 
the months following marriage. At
first sight this may seem an unlikely proposition but it
was shown that the existence of such a mortality spike 
is indeed supported by statistical evidence. (Richmond et al 2016b)
\qpar

There is a noteworthy difference between widowhood and
marriage in the sense that widowhood occurs usually in old
age when there is already a situation of frailty. A similar case
considered
in the papers mentioned above concerns elderly persons who
move from their home to a nursing home.  The few 
data that are available for such cases
often display a mortality shock in the months 
following admission. Here too, there is certainly a frailty effect.
The biological phenomenon considered
in the present paper will also involve these two components:
(i) a shock in the form
of an exogenous stress factor and (ii) a situation of frailty.

\qA{Births seen as transient shocks}

Coming now to biological cases, in humans and in many
other species, birth constitutes a major
shock which, not surprisingly, results in
a huge mortality spike (Berrut et al. 2016).
This means that the death
rate in the days following birth is much higher
than the death rate {\it in utero} in the weeks preceding
birth and also higher than the neonatal death rate (i.e.
average death rate over the first 6 weeks).
It can be observed that a large part (over
50\%) of these early deaths are due to congenital
malformations which were compatible with life
{\it in utero},  but become lethal in the new
environment; lung defects are a clear example.
The congenital
malformations may appear during embryogenesis
but may also be triggered by defects 
already present
in the initial  female and male germ cells%
\qfoot{The extent of such defects is attested by
the huge mortality peak in the initial phase
of the embryogenesis (see Chen et al. 2020)}%
,
i.e. the oocytes and spermatozoons.

\qA{How the process of fast cell replication induces frailty}

The question that we address in this contribution is whether or not
cells experience a transient shock (similar to birth) in the 
process of division. We will propose reasons valid for both eukaryotic cells
and bacteria  for a state of fragility associated with cell division.
\qpar

For a bacteria like {\it Escherichia coli}, the generation time
is about 20mn. During this short time, many complicated
processes have to be carried out (Fig.1). One can keep in mind
(Soufi et al. 2015)
that the proteome (i.e. the entire set of proteins expressed
in the cell) of {\it E. coli} is comprised of some 2,300 different
proteins with concentrations ranging from just a few copies to some
300,000 for the more abundant.  Appendix A gives an
exemple of the working of a protein enzyme.

%
 \begin{figure}[htb]
 \centerline{\psfig{width=15cm,figure=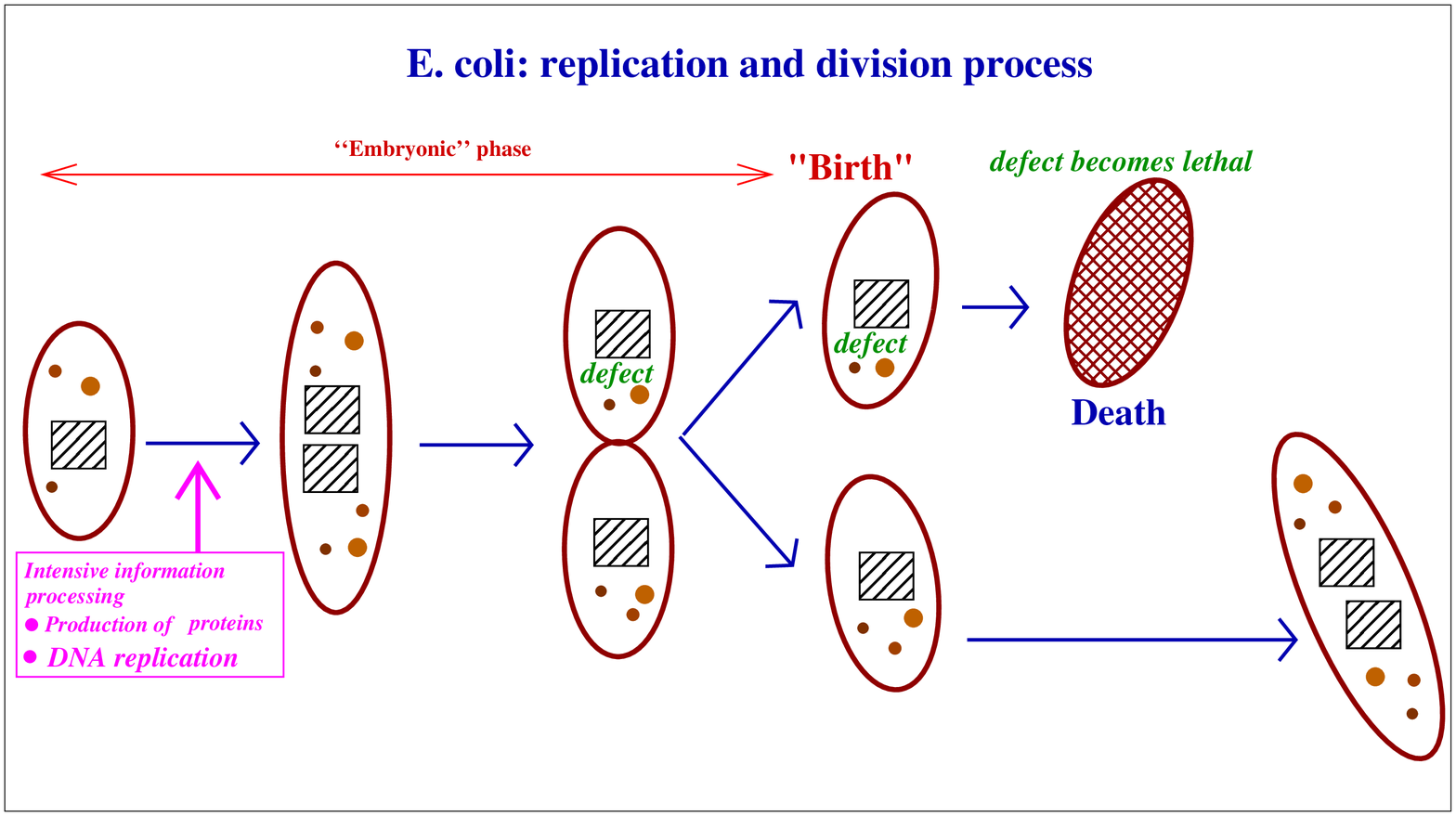}}
 \qleg{Fig.1 \quad Replication and division of {\it E. coli}.}
 {The first phase is marked by intensive protein production
and  replication of the genome and other components.
It is the analog of the interphase in eukaryote cells.
The
replication of all the components of the cell is certainly
a major task of information processing.
On the right-hand side of the diagram it was assumed  that,
due to a defect in the replication procedure,
a key-protein
$ P_1 $ was not produced in appropriate amount in the upper cell.
Despite that, the situation 
remained manageable as long as exchanges were possilble
with the mother cell.
After separation the lack of $ P_1 $ may prove
lethal. This fragility induced by division will be called
the ``replication frailty''.}
 {}
\end{figure}
%

In the manufacturing of an airliner accuracy requirements differ
from one component to the other. If millimetric accuracy may be
acceptable for the airframe, there are tighter requirements for
the engines. 
There are good reasons to think that in the process of Fig.1
the most critical part is the replication phase.
\qbu Any fault in the production of the proteins which function
as the bricks and tools of the replication process may subsequently 
lead to fatal failures of the kind shown in the figure.
\qbu It is true that there are control and reparation mechanisms.
For instance if a gap appears in the replication of the DNA the two
ends may be joined again thanks to the cohesion forces which exist
between consecutive elements. The problem is that control and 
reparations take time whereas in fact the whole process is designed
to achieve high speed. For instance, instead of starting from
one location of the DNA molecule, replication starts simultaneously
from different locations%
\qfoot{More precisely, (i) the replication starts
in two directions  and (ii) once the specific 
starting point has been duplicated, replication may also start
from this point.}%
This speed requirement is a major 
cause of fragility. It will be seen shortly that bacteria having
a doubling time of 20mn are much more vulnerable 
to stress factors than others
whose doubling time is 5 hours.
\qpar

Another crucial moment is the separation.
However, what makes it risky  are the
mishaps that occured in the replication.
To work well,
{\it E. coli} cells need thousands of proteins. Now,
suppose that something went wrong in the replication
with the result that an essential enzyme-protein $ P_1 $ is not
produced in appropriate quantity. As long as exchanges
remain possible, the daughter cell can rely
on the $ P_1 $ produced by the mother cell. This
kind of sharing becomes impossible after 
separation. In other words, the 
death rate of daughter cells will reflect the 
replication defects, much in the same way as post-birth
human mortality reflects the manufacturing anomalies
that occurred during embryogenesis. Naturally, not all
defects are lethal immediately after separation. 
In humans it is well known that
some congenital defects (e.g. heart valve anomalies)
become life-threatening only in old age when compounded
with other factors, e.g. increased membrane rigidity.
Whenever they occur, non lethal
replication anomalies 
leave daughter cells in a state of fragility.
If in addition the cells are subjected to adverse
environmental conditions (e.g. X-rays, non optimal temperature,
pressure or acidity) this fragility will
bring about a death rate spike.

\qA{How the paper proceeds}

So far we have explained
the mechanisms on which our understanding relies.
In the rest of the paper we adress two questions.
\qee{1} How compatible are the data on death rates in 
bacteria with the existence of a transient state of frailty  during
cell division.
Many experimental results are available which
describe the response of bacteria to harmful
exogenous factors. It appears that the populations with high
growth rates are much more sensitive to those harmful conditions
than those with slow growth rates.
Illustrative evidence will be presented and discussed.
\qee{2} Has the transient shock model predictive
implications not yet considered and which may
suggest new experimental tests?
Can the tests on bacteria be extended to other systems of growing 
organisms, e.g. replication of viruses in cells, replication of
eukaryote cells, cleavage of first cell in an embryo?

\qI{Response of bacteria to adverse exogenous factors}

Adverse exogenous factors can be of many kinds:
\qbu Ionizing radiations (e.g. X or gamma rays)
\qbu Low or high  temperature 
\qbu High pressure
\qbu Inadequate acidity
\qbu Disinfectants and antibiotics
\qbu Pulsed electromagnetic fields
\qbu Ultra violet light
\qpar

Here we wish to focus on parameters which modify the
division rate. The following can be considered.
\qbu Normal cells versus cancer cells will be considered in the next
subsection. As is well known cancer cells are more vulnerable to
X-rays than normal cells.
\qbu Different strains of bacteria have generation times (also
referred to as doubling times) ranging from 20mn to several
hours. Those with the fastest growth are the most vulnerable.
An illustrative example will be given.
\qbu Any new cultures of bacteria starts with a period of fast exponential
growth but after a while it enters a stationary phase in which 
its birth rate converges toward zero.
It will be seen that bacteria taken from the rapid growth regime
are much more vulnerable to stress factors than those from
the stationary phase. An illustrative example will be described.
\qpar

Among the many publications on these effects 
a good introduction is
provided by a series of three papers 
which emphasize the role of the growth rate, namely
Ihssen et al. 2004, Berney et al. 2006, Lindqvist et al. 2014.
They consider the cases of temperature jumps, 
ultra violet light and  acidity, 
The impact of very high pressures of over 1,000 bars
has been studied in Benito et al. 1999 and Manas et al. 2004.

\qA{Early observation of the division effect}

One of the first observations of the replication effect was made in
the early 20th century when it was discovered
that ionizing radiations can destroy cancer cells
without too much affecting surrounding cells.
After coming about through
the activation of germ-like stem cells,
normal cells multiply
exponentially until the required number is reached.
Then, they enter
a stationary phase in which their growth rate 
is almost zero. On the contrary, cancer cells can grow
exponentially without limitation.  As an example, 
in the case of small
cell lung cancer  the doubling time of the cells can be 
as short as 30 days.
\qpar

Why were cancer cells more affected than normal cells?
There are (at least) three possible interpretations.
\qee{1} One explanation is to say that the radiations
disrupt the DNA to the point of making any subsequent
division impossible. Yet, taken alone, this reason
cannot explain why normal cells are less affected.
\qee{2} Another explanation is to say that in the 
stationary phase the cells
develop a number of protections which shields them
from the radiations. 
\qee{3} The third explanation relies on the fact that,
as decribed above, cell divisions bring about a state
of fragility which makes them vulnerable, not just
to radiations but to
any adverse factor. In this interpretation
the more divisions cells
undergo, the more vulnerable they are.
\qpar

One great difference between interpretations
(2) and (3) is that (3) not only explains the radiation
effect but has also predictive implications for cell 
responses to other harmful factors. Actually, (2) and (3)
do not exclude one another but are rather complementary.
It is true that in the stationary phase cells develop
means for long term preservation. The transformation
of cells into spores is a clear illustration of this
kind of mechanism. Yet, explanation (2)
has no predictive power; it gives a description of the
sporulation  process whenever there is one, 
but it cannot predict what cells 
have the ability to form spores or what other preservation
mechanisms can be activated.
\qpar

Quite surprisingly, despite its early observation, 
the fragility mechanism has been all but forgotten
in subsequent decades. A 343-page manual on how 
bacteria are affected by radiations,
published in 1973 by the ``International Atomic Energy
Agency'' (IAEA) does not even mention this effect. 
In contrast,
There have been numerous papers
describing various protection mechanisms of type (2). 
As
such mechanisms are species-dependent, it may be 
a case of trees concealing the forest.  
We hope that by  highlighting the frailty explanation the present
paper will bring it back to the attention of researchers.

\qA{Prokaryotic versus eukaryotic cells}

Why should replication and division put bacteria
(that is to say prokaryotic cells)
like {\it E. coli} in a particularly vulnerable situation?
The fact that in bacteria the replication 
of the DNA progresses at the impressive speed
of 1,000 nucleotides
per second, compared to 100 per second for eukaryotes
suggests that such high speed processes are probably
hypersensitive even to small perturbations.
In the present paper we will not discuss this point
in more detail; our plan is to accept it as a fact and 
analyze its implications. 
\qpar

Radiotherapy relies on the replication
effect in eukaryotic cells and this makes us expect an even
stronger division effect in bacteria.
In the following subsections we give examples which
quantify this effect. It will be shown to what extent
the death rate occasioned by adverse conditions
is conditioned by the frequency of replications.

\qA{Effect of ionizing radiations}

We start with this case because, as described above, 
it was historically the first to attract the attention 
of researchers%
\qfoot{It has the additional advantage that radiations
are {\it always} harmful. In contrast, for assessing the effect
of a surge in temperature, one needs to know what is
the cells'  optimal temperature.}%
. 
\qpar

The following observation conveys very clearly the idea of the
frailty effect. \qL
It has been shown 
(see Lacassagne  1930 and Lea 1947, p.307) that when a cell
is irradiated death does not occur immediately but at (or following) the 
next division. For instance, 10,000 roentgens delivered to yeast
allows the  cells to divide but their daughter cells are usually
unable to divide  further and eventually die. A much larger dose
is required to kill the cells immediately; thus, a dose of 30,000
roentgens will kill only 50\%. 
\qL
The same effect can be seen for
eukaryote cells. When chick tissue growing in culture is 
irradiated,  a dose of 100 r suffices to cause the death of the cells
which attempt to divide,  but 2,500 r are required to cause the death
of an appreciable proportion of resting cells, i.e. those which do not
divide.
\qpar

In an experiment described in the aforementioned 
report  of IAEA (1973, p.41)
different populations of bacteria
were subjected to increasing doses of radiations until
only 0.1\% survived (Fig.2).
This was done successively with
two species. For {\it Micrococcus radiodurans}  
($ G=5 $h)
this result is obtained with a dose of 1.20 Megarad
(rad means ``radiation absorbed dose'', this old
unit is related to the SI unit called Gray by the
relation: 1 Gray=100 rad).
%
 \begin{figure}[htb]
 \centerline{\psfig{width=9cm,figure=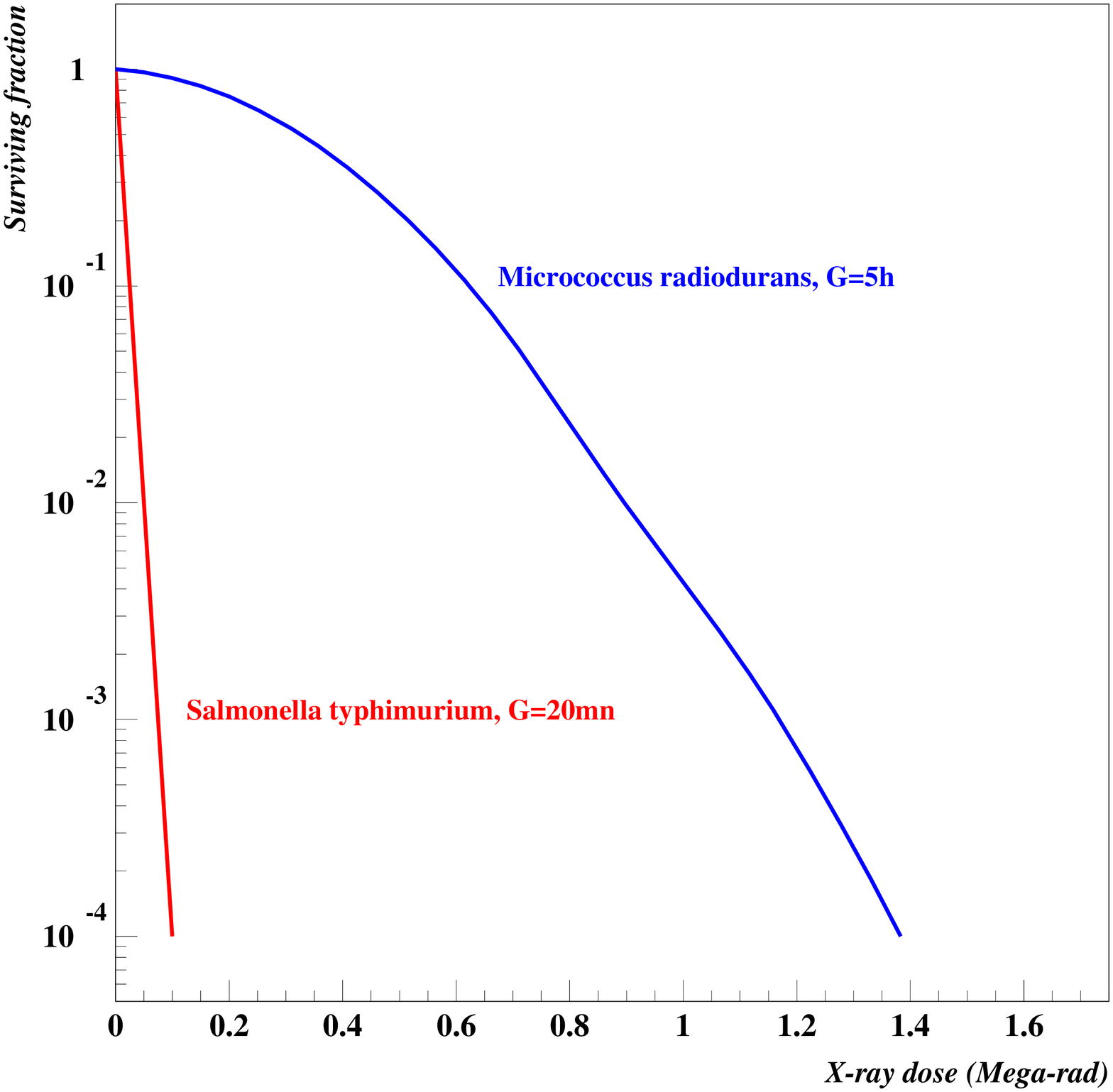}}
 \qleg{Fig.2 \quad Sensitivity of bacteria to X-rays.}
 {The sensitivity to X-ray radiations is strongly conditioned
by the generation time G of the bacterial strain. 
The faster the bacteria divide, the more sensitive they are 
to radiation.}
{Source: The data are from IAEA (1973,p.41)}
\end{figure}
%
In contrast, for {\it Salmonella typhimurium} ($ G =20 $mn), only
0.05 Megarad were required to get the same result.
In other words, division of $ G  $ by 15 resulted
in a reduction of the dose by 24.
\qpar

An alternative approach (see Appendix B)  could be conducted
by comparing the required dose in the exponential phase
of {\it M. radiodurans} to the dose for the same bacteria
but in the stationary phase (Serianni et al. 1968, p.198).
One finds that a dose about 3 times larger is required
in the stationary phase. In short, the faster the cells divide the
more sensitive they are to radiation.

\qA{Effect of cold}

Among numerous papers devoted to studying the effect of
stress factors, 
the experiments done by James Sherman and William Albus (1923) are of
particular interest for (at least) two reasons.
\qbu Whereas most papers study the effect of temperature rises,
this one investigates the effect of cold.
\qbu In contrast with many papers in which the protocol is neither
well designed nor well explained (see Appendix B), 
here the methodology is carefully designed and well explained. 
\qpar

The authors start with the observation that it is well known that a
brief exposure to cold (above the freezing point of the medium)
causes almost no reduction in the number of viable {\it E. coli}
bacteria
as counted by the CFU (Colony Forming Unit, see Appendix B)
method, provided that the population is in the stationary phase.
Then, the authors describe the following experiment. 
\qdec{{\bf First experiment.}
A culture of {\it E. coli} in 1\% pepton  was kept for 8 days
at laboratory temperature (25 degree Celsius) before being 
diluted in distilled water
at a temperature of 2C. The numbers of viable cells
were counted on a small sample, immediatly after dilution ( $ N_0 $)
and again after one hour exposure at this temperature ($ N_1 $).
They found (average of two trials):
 $$ \hbox{Old culture (8 days)} \Longrightarrow N_1/N_0=0.99\pm 0.014 $$
}

Then, they repeated the same experiment with a culture which was only
4 hours old. This time they found the following numbers (again average
of two experiments):
$$  \hbox{Young culture (4 hours)} \Longrightarrow N_1/N_0=0.47\pm 0.26 $$

They were tempted to conclude
that there was an over-mortality among the young
cells due to sudden cooling.
However,  the authors observe quite judiciously that it was not
possible to conclude that it was the cooling which caused the
mortality
for in fact there were {\it two} changes: (i) temperature and
(ii)  transfer from a nutrient medium to distilled water.  At this point the
authors could have performed two tests: 
\qee{1} Replace the water at 2C by water at 25C. This would show
the water effect alone.
\qee{2} Replace the water at 2C by pepton medium at 2C. This would
show the temperature effect alone.
\qpar

The authors decided to do the second test. In a sense this was a
logical decision because it is in the temperature effect that they were
interested. However, just as a confirmation, it would also have been 
interesting to try the first test (see Appendix C).
\qpar

\qdec{{\bf Second experiment.}
A culture of {\it E. coli} in 1\% pepton  was kept for 12 days
at laboratory temperature (25C) before being 
diluted in pepton medium
at a temperature of 2C. The numbers of viable cells
were counted immediatly after dilution ( $ N_0 $),
and after 1 hour exposure at this temperature ($ N_1 $).
They found (each measurement was done only once):
 $$ \hbox{Old culture (12 days)} \Longrightarrow N_1/N_0=0.96 $$
$$  \hbox{Young culture  (3 hours)} \Longrightarrow N_1/N_0=0.42 $$
}

The fact that the $ N_1/N_0 $ ratio for the young culture was almost
the  same as in the first experiment suggests that the pepton to
water (both at 25C) transition had almost no effect in terms of
mortality. Naturally, in distilled water a cell cannot produce
any daughter cell because the basic materials (e.g. carbon) are unavailable.
\qpar

So, is this experiment the ultimate test?
Not entirely. From the data given by the authors  we observe
that in the old culture, $ N_0=970 $ per cubic cm, whereas in the
young culture: $ N_0=3,420 $ per cubic cm.  In other words, the density
of bacteria in the young solution is over 3 times higher than the
density in the old culture. Can that be of importance? Probably not, but
to be on the
safe side it would have been easy to dilute the young culture so as to
reach the same density as for the old culture. 
\qpar

The authors measured also the number of cells after 2 and 3 hours at
2C. There were only slight decreases:
 $$ N_2/N_1=0.80 \quad N_3/N_2=0.85 $$

These numbers are consistent with the interpretation that the effect
is mostly due to the cells which were in the process of division
shortly before cooling.
Such cells were very numerous in the young culture at the moment
of its dilution in the medium at 2C. What happened to them?
Their growth was certainly stopped. 
Were they killed or only inactivated in the sense of becoming
unable to form colonies? We do not really know.
\qpar

Novadays, measurements with a spectrophotometer would allow us 
to know whether or not some cells could successfully complete their
division in the minutes after cooling. One would not be surprised that
cells in the final phase might be able to terminate
their division. 

\qA{Effect of other stress factors}

It can be noted that in the previous experiment 
by Sherman et al. (1923)
the replication effect was rather weak in the
sense that the number of viable cells was divided by only 
two whereas
in the X-ray experiment we have seen much greater reductions.
This is understandable because the 
cooling froze all new divisions; only the few cells whose
division process was well advanced were affected.  \qL
This interpretation
is confirmed by the next experiment in the same paper.
In this case, the stress factor consisted in exposure in
a solution of 2\% NaCl during 1 hour. This factor 
stressed the cells but did not stop the divisions. As a result,
the number of young cells was divided by 30. \qL
In another experiment (in the same paper) the stress factor was 
exposure to a temperature of 53C during 20mn. In this case, the
excess reduction of the young cells with respect to the old
cells was a factor 75.

\qA{Internal changes when exponential growth slows down}

We have already observed that there are internal changes.
How important are they? In this respect
the comparison between {\it M. radiodurans}  
and {\it S. typhimurium}, both of them considered in 
their exponential phase,
has a clearer significance than experiments of type (2). defined
in Appendix B.
Why?\qL
When a culture gradually moves from the exponential phase
to the stationary phase, many internal transformations take place.
One of these, which is mentioned in the literature in
connection with cell response to various stress factors, is the increase
in the RpoS factor%
\qfoot{Although irrelevant for our argument, the origin
of this cryptic name can be explained as follows.
R stands for the first letter of RNA; po
stands for polymerase, a catalyst which produces chained
molecules and is used in DNA to RNA transcription; finally
S refers to the fact that this factor is only active in the
initiation of this transcription. Another often 
used protective technique against 
radiation and other stress factors is clustering .}%
.
The production of this factor increases as the population
enters the stationary phase and we are told that it may
also be activated by a situation of stress. However,
this can by no means explain the difference observed between 
{\it M. radiodurans}  and {\it S. typhimurium} for
we are talking here of bacteria in their exponential phase.
In contrast,
the explanation by vulnerability during divisions
remains valid. In short, although we do not deny that 
there may be 
some specific defense mechanisms, we think that,
because of its broad validity, the replication-frailty effect
should be considered first.
\qpar

\count101=0  \ifnum\count101=1

\qA{Effect of acidity}

The case which is documented in Fig.2 is of particular
interest for our purpose. Why? \qL
In the next section we will propose as a conjecture
that the relationship between the doubling time and the
death rate can be extended to cases where there is no
specific exogenous stress factor whatsoever. Therefore, a case 
with light stress provides a stepping stone to
stressless cases.
\qpar

%
 \begin{figure}[htb]
 \centerline{\psfig{width=8cm,figure=acid.eps}}
 \qleg{Fig.2 \quad Sensitivity of E. coli to acidity (pH=4.6).}
 {The curves show the component due to the division effect.
The upper curve (open squares, growth rate=zero) would
correspond to the stationary phase of a batch growth culture.
The proportion of survivors was determined through
the standard technique of counting colony forming
units (CFU). Although this technique has a fairly low
accuracy, here it is perfectly acceptable because one
measures very large changes. As the experiment lasts
4 days it can only be done in a chemostat, that is to
say a culture which takes place in a stationary flow. }
{Source: The experimental data are from Lindqvist
et al. 2014.}
\end{figure}
\fi

\qI{Conjecture and predictive implications}

\qA{Statement of the conjecture}

We have shown that
instead of relying on a multitude of protective mechanisms
which are strain- and stress-dependent, all the 
experimental evidence
about responses to stress can be explained by assuming
that cell divisions make the cells particularly vulnerable to
any pre-existing stress factor. Is it possible to extend
this mechanism to situations where there is no specific
stress factor? We believe so. 
This conjecture can be stated
as follows.
\qdec{\it {\color{blue} Frailty conjecture}: 
The death rate of a culture of bacteria grown under
nearly ideal conditions should be correlated
to its growth rate. Consequently, it
should be lower in the stationary phase than in
the exponential phase.}
\qpar

As all conjectures, this one relies on several
assumptions. For instance, it is assumed that the natural stress
factor due to non-perfect conditions
does not increase as the growth rate of the culture
decreases.

\qA{Supportive argument}

A simple  argument goes as follows. \qL
A cell culture
is never in a situation that is 100\% ideal. Nutrients, 
pH, supply of dissolved oxygen and so many other parameters
are never all completely optimum for the simple reason that in 
fact we do not know exactly what the most perfect conditions 
are.
The discrepancy between existing and perfect conditions are
seen by the bacteria as a light form of stress.
Thus, what we have said about stress factors leads
to the prediction that the natural death rate
(usually denoted  by: $ \mu_n $) will be lower in the
stationary phase than in the exponential phase.
\qpar

It can be observed that the previous
prediction is rather counter-intuitive. Why?
Usually, the stationary phase is seen as resulting from
adverse conditions in the form of insufficient nutrients 
and accumulation of residues. There is no doubt that 
such unfavorable conditions do exist. This means that in a batch 
experiment the two effects occur simultaneously and
are in competition with one another.
In the early stationary phase, the growth rate effect will
probably dominate but it is certain
that after a while the effect of bad conditions will prevail.
\qpar

A graph published in Wilson (1921, p.430) suggests that
for a population without external stress factor,
if the stationary phase starts around time $ T_s $ 
(in the case presented in the paper $ T_s=5 $h),
then the death rate begins to swell
substantially after time $ 2T_s $ and the number
of viable cells falls back to its initial level toward time
$ 4T_s $.
\qpar

One can get rid of these unwanted competing effects by using 
a chemostat in which nutrients
are renewed and residues eliminated (see Appendix B).

\qA{Experimental tests of the conjecture}

Although the previous argument seems reasonable we do not put
too much faith in it. The real test will come from
experiments. 
\qpar

However, one must realize that the required
measurements will not be easy. Why? \qL
When there is a stress factor, the population 
falls very substantially, for instance
it may be divided by two or three within one hour.
Such big changes can easily be assessed even with
fairly crude techniques such as the CFU counting technique.\qL
On the contrary, for the natural death rate of 
{\it E. coli} in the exponential phase, one expects a death
rate of the order of 1 per
1,000 living cells and per hour. 
If our conjecture is correct the death rate
in the early stationary phase will be even lower.
Needless to say, to get reasonable estimates
for such low death rates 
one must be able to measure the number of dead
cells with high accuracy. 

\qA{Amplification of small differences in growth rates}

Usually,
growth rates are not easy to measure. The replication effect
opens the possibility of replacing growth rate measurements
by death rate measurements. 
\qpar

Translation of  growth rate differences  into death rate differences
can be done by applying a  standardized stress factor  (e.g. X-rays or 
ultra-violet light).
This may be useful  in fields (e.g. oncology) where differences 
in growth rates are important.

\count101=0  \ifnum\count101=1
%
 \begin{figure}[htb]
 \centerline{\psfig{width=15cm,figure=acid.eps}}
 \qleg{Fig.3a: Changes of the growth rate $ \alpha $
as a population moves from exponential to stationary phase.
Fig.3b: Effect of $ \alpha $ changes on sensitivity
in response to a temporary 48C temperature jump.} 
{The growth rate $ \alpha $ is the standard exponent:
$ y(t)/y_0=\exp(\alpha t) $ where $ y $ denotes
the size of the population at time $ t $. If one excepts
the transient initial growth, $ \alpha $ decreases 
steadily from its top value (which corresponds to a
doubling time of 24mn) to zero. The three straight lines
marked 3,2 and 1 hour show the response to shocks of
decreasing strength. Even the mildest of them substantially
affects survival. The fact that the stress brings about
a fall in the population
means that the death rate due
the stress surpasses the natural growth rate.}
{Sources: Data from Lindqvist et al. (2014) and
own experiments.}
\end{figure}
\fi
%

\qI{Extensions to embryo and viruses}

``Is it possible to express our conclusions concerning the effects 
that we have described in such a way that some general
principles emerge from them?''
\qpar

This is the question raised in one of his papers on this
topic  by Nobel prize winner Andr\'e Lwoff (1959, p.120). 
To this end we state the replication effect
in more precise form and then we will ask if its presumed
applicability can be extended. 
This will make the testing game
more risky but also more rewarding.
It would be pointless to propose an explanation which cannot
be disproved.

\qA{Enlarging and focusing}

Whether for prokaryotes or eukaryotes, the division process 
goes through
similar successives steps. Given that the overall objective is the 
same, it is
hardly surprising that there are common features: production of
a great number of proteins, replication of the DNA and transcription of
the RNA, spatial separation of mother and daughter  components 
at the two ends of the cell and finally separation%
\qfoot{Although the broad mechanisms are fairly clear many questions
regarding  their real implementation remain unanswered.}%
Actually, such
steps are indispensable in all cases where a mother cell gives rise
to a daughter cell. Apart from the two cases already considered, one
can add to them two others not yet considered.
\qee{i} The initial cleavage of
a zygote, that is to say the first step  in the process that will lead
to an embryo and eventually to the birth of a new individual. 
\qee{ii} The replication of viruses in their host cell.
\qpar

Having thus enlarged the question, we wish at the same time 
to make it more focused
by asking what is the most crucial part of the division
process as far as manufactoring mishaps are concerned.
In fact, this question was already addressed earlier (see also 
Appendix A)
In the framework of our paper, it is clearly the early phase of the
division which is considered to be the most critical. Called
interphase
in the case of eukaryote cells, this phase (more precisely
 the G1-S phases of the interphase)
comprises the synthesis
of many proteins and the replication of the DNA. Our conjecture is 
that it is during this phase that the cell is the most vulnerable
to external perturbations. The main question is whether the
mechanism brought to light for bacteria and eukaryote cells remain
valid for viruses and embryonic development. It will be addressed in
the following subsections. However, our purpose is only to open new
perspectives that can be 
considered more fully in subsequent studies.

\qA{Frailty mechanism for viruses}

Broadly speaking the existence of such a mechanism is both
self-evident and yet still mysterious. It is self-evident 
because
viruses are living entities only during the time when they 
get replicated inside an host cell. Naturally that does not mean 
that they cannot be destroyed outside of their host-cells.
Any molecule, and particularly macromolecules, can be broken 
up into pieces by radiation or excessive temperature%
\qfoot{Lwoff addresses this point by saying that mere ``thermal
inactivation of viral particles [i.e. outside of the host 
cell] in  few hours in temperatures ranging from 37C to 41C
is negligeable.''}%
.
\qpar

Once inside their host-cells the replication of the viruses happens
to be highly temperature dependent (see Lwoff 1959, p. 111). 
Unlike chemical reactions which, according to Arrhenius' law,
become faster when the temperature is (moderately) increased,
for viruses it is the opposite. As soon as the temperature 
becomes higher than their optimum temperature, their replication
is reduced by several orders of magnitude. Lwoff gives measurements
for the poliovirus which show that at 40C their development
is reduced by a factor of almost 1,000 with respect to
what it is at 37C. Such a high vulnerability to a fairly
small temperature change is indeed coherent with the 
high growth rates achieved by some viruses. Here again,  the fastest
the division, the higher the vulnerability to stress factors.

\qA{Frailty mechanism in embryogenesis}

The early steps of embryogenesis consist in a  replication
process. Here we have a limited goal. 
We wish to draw attention on just one point 
which may help us to better understand the bacteria
case.
\qpar

For most species of fish the fertilization of the eggs
produced by the female occurs outside of the body 
which means that it
can easily be observed from the very first moment on. 
In this way one
can measure the death rate of the embryos during their whole
development. It appears that the bulk of the deaths
is concentrated toward the beginning of the development
and that it can reach very  substantial rates of the order 
of 10\% (see Chen et al. 2020). This is 100 times more than
the death rate for {\it E. col}i cells during the phase 
of exponential growth. Why do we have such high rates here?
We believe there is a simple reason which is related to how the 
female and male germ cells are produced. 
\qpar

In the case of {\it E.coli} any cell can be considered
as an initial cell that
has been produced by a mother cell through a
standard process of
division that is known to work fairly well
with about only one defect in 1,000 divisions.
For the oocytes produced by the female it is a very different 
matter. They are produced through an intricate process which
comprises many steps (including two divisions).
Although there may be some control mechanisms 
along this production 
process there is no screening process that would eliminate
the defective eggs. In fact, the actual screening takes place
in the first divisions of the embryogenesis%
\qfoot{We did note mention possible defects 
of the sperm cell because it is about 1,000 times
smaller than the egg, nevertheless defects cannot be
excluded altogether.}%
.
\qpar

If a parallel can be helpful
it would be with the launch of a rocket compared
to a departing train in a station.
The departure procedure (e.g. announcing imminent
departure on the train's loud speakers,
closing the door, starting the engine)
has been tested multiple times in
previous stations. On the
contrary, although the individual components
(e.g. the engine, the navigation system)
may have been controlled in many
ways, the launch will be the first time for
all components to work together. It is well known that 
despite multifold tests, the probability of failure
remains usually of the order of 5\%. Similarly, the division
of the zygote is for the female and male
components the first time to work together.

\qI{Perspectives: complementary experimental tests}

In this section we describe a number of open questions.
Experiments are proposed which should help to answer them.
\qpar

Does cell concentration constitute a stress factor for a 
sample of cells?
This is an important question that we did not consider so far.
Two interrogations come to mind immediately.
\qbu The term ``concentration'' can be understood in two ways:
(i) with respect to the total volume of the liquid, $ c_v $ 
(ii) with respect to the amount of nutritive elements, $ c_f $,
contained in the liquid. For the sake of brevity we call them
``volume-concentration'' and ``food-concentration'', respectively.
When a culture is growing, both $ c_v $
and $ c_f $ decrease. It would be of interest to know the effect
of each factor separately. Clearly, if $ c_f \rightarrow 0 $, division
will stop but 
we do not really know what happens when 
$ c_v \rightarrow 0 $ while $ c_f $ is kept constant.  
The experiment summarized in Fig.3 should answer this question.
\qbu The term ``stress factor'' can be understood in two ways: 
(i) a factor limiting growth (ii) A factor inducing higher mortality.
For the sake of brevity
these meanings will be referred to as  ``growth stress factor'' and
``life stress factor'' respectively.  \qL
In the papers reviewed above
the reduction in the number of viable cells
was well documented but there was little information about
birth rate changes. Probably, it was assumed that a fall in
viable cells would necessarily be paralleled by an interruption
of the division process. However, that is by no means evident.
Cases of human populations having high death rates coupled with
high birth rates are well documented in developing countries.

\qA{Is the volume-concentration of cells a growth-limiting stress
  factor?}

The experiment consists in three steps.
\qee{1} After the two samples have been extracted from
the culture their optical density and their
their doubling time are measured.
\qee{2} Dead cells are added to sample $ A $ and the same
volume of medium but without cells is added to sample $ B $.
%
 \begin{figure}[htb]
 \centerline{\psfig{width=9cm,figure=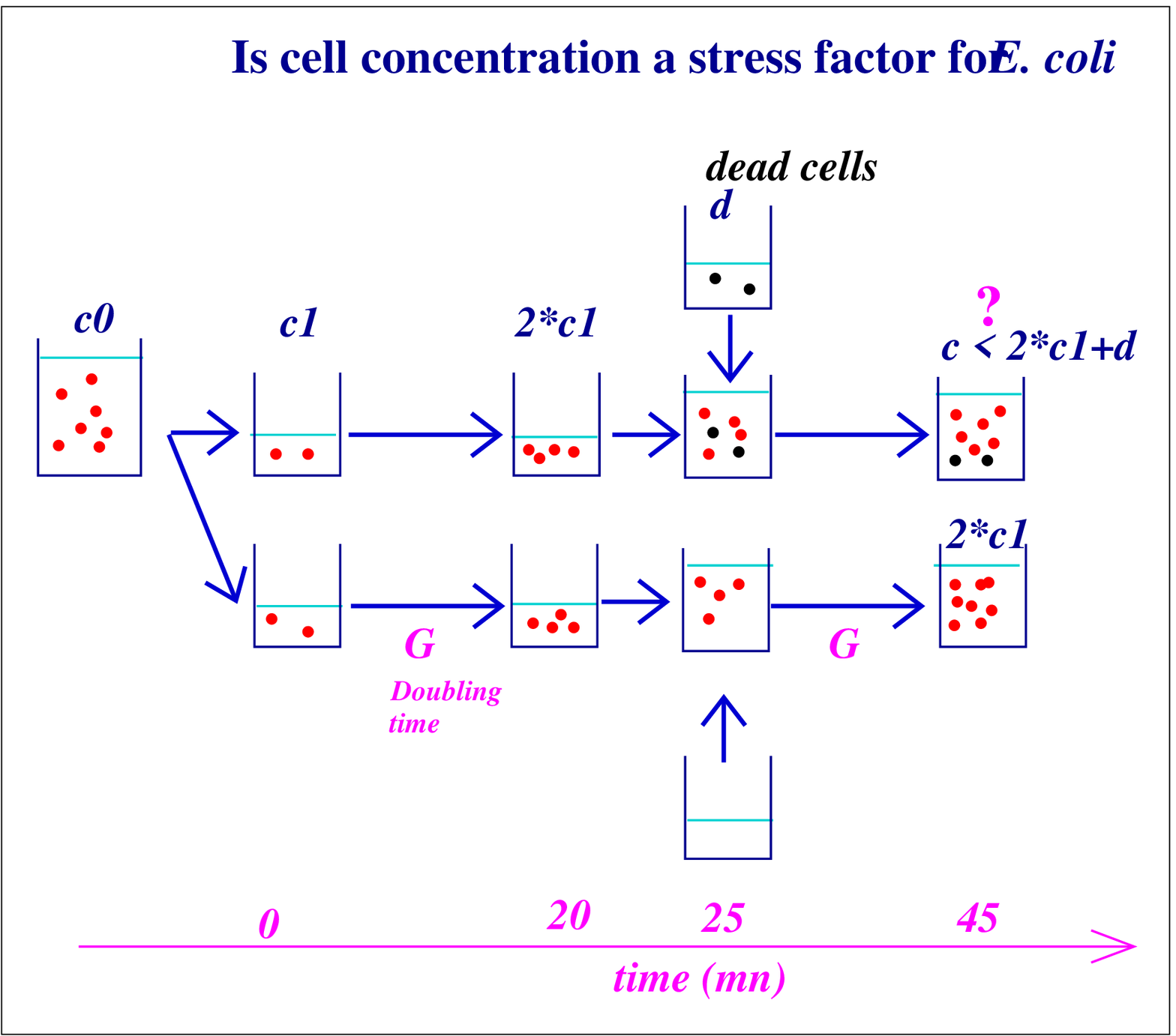}}
 \qleg{Fig.3 \quad Sensitivity of bacteria to volume concentration.}
 {In order to change the volume concentration without changing
the food concentration dead cells are added to the culture.}
{}
\end{figure}
%
\qee{3} The doubling time
is measured again in order to see
if (and by what amount)  the growth of sample $ A $ was
slowed down with respect to the control sample $ B $.

\qA{Effect of: medium $ \rightarrow $ water transition.}

In the analysis of the experiments conducted by Sherman and 
Albus (1923), we came across the question of whether it is 
possible to define what can be called
the momentum of inertia of a culture.

 \begin{figure}[h]
 \centerline{\psfig{width=5cm,figure=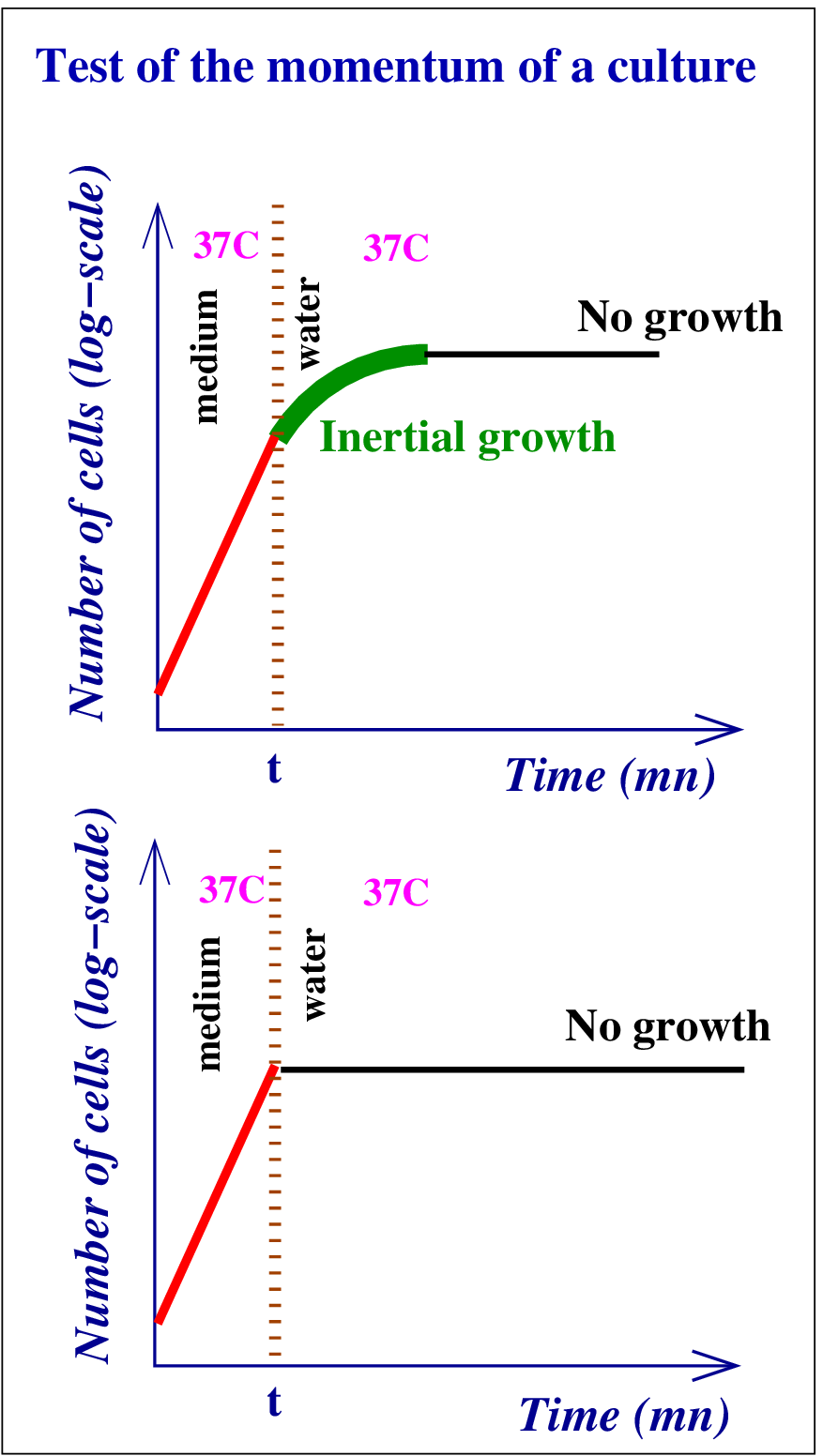}}
 \qleg{Fig.4: What amount of building materials
are stored in the cells during exponential growth?}
{The graph at the top would be observed
in case of an important stock of advance material.
On the contrary, the bottom graph corresponds to 
the assumption of ``just in time'' production that is to 
say with the minimum of advance storage.}
{}
\end{figure}

Simply stated,
this notion gives an estimate of how long a culture can
continue to grow once it is put in distilled water. 
Here, we propose a simple experiment .
\qpar

Two different growth surges can be expected.
\qee{1} The growth based  on ``endogenous'' resources may 
occur shortly after
the transfer into water. One knows that in the phase of rapid
growth of {\it E. coli} the replication of the DNA destined to the 
{\it grand-daughter} starts before the separation of the daughter 
has occurred. This means that the cells certainly contain more
materials than the amount strictly needed to produce a single daughter. 
\qee{2} An exogenous resource is provided by dead cells
which release some of their components.  This is a fairly 
obvious effect in which we are not really interested.
One hopes that the time lag between the first and second effects
is large enough to allow a clear distinction.
\qpar

This simple experiment should be able to tell us what is the amount of
``building materials''  that are stored in the cell at different
moments of the growth cycle.
One would expect that
this amount dwindles when the speed of growth decreases.
\qpar

This effect can also be used to probe the age distribution of
the population. To make this point clearer let us assume that the
population is synchronized. Then, if all cells are freshly divided
they do not possess the materials necessary for replication and
therefore will be unable to divide. Thus, one expects the bottom
graph in Fig.4. On the contrary if all cells are mature and
about to divide one expects the top graph in Fig.4. \qL
In a mixed case, the length of the inertia phase will give 
an estimate of the proportion of mature cells.

\appendix

\qI{Appendix A:  Manufacturing defects versus mutations}

In this appendix we wish to explain the implications
of a defective blue-print  (as a result of mutations in
the genome) as contrasted with mishaps in how the prescribed
tasks are carried out.
It is true that mutations of the genome
may be increased by external factors
such as X-rays or ultra violet light. However, even
if the blue-print
is correct there will be 
defects in the implementation process. This question is outside the
scope of our paper but it is important to realize that mutations of 
the genome are not the only mechanism which leads to anomalies.

\qA{Defects and mishaps}

Instead of the expression ``manufacturing defects'' we were
at first
tempted to use more specifically biological expressions such as
``error of metabolism'' or ``enzymatic defects'' but then we realized
that in fact we have in mind few clear cases of enzymatic defects that
would illustrate our argument. In contrast, examples of
manufacturing defects can be cited easily. Consider for instance
the construction of a building. Naturally it will proceed
in accordance with the blueprint established by an architect.
But in its implementation there may be many mishaps. In hot
weather concrete may dry too fast thus impairing the mechanical
properties of the walls. Freezing weather may also cause
problems. In other words, at each step, external
non-optimal conditions may exist. In some cases the effect may
be light while in others they may lead to structural flaws.
\qpar

In biology inappropriate temperature
or pH conditions may lead to
inordinate protein production: too little, not at the right location
or not in the appropriate 3-d configuration (for instance a cis-
instead of a trans-molecular structure), any of these defects
 may possibly lead to life threatening abnormalities.
 Naturally external parameters are never 100\% optimum,  but may 
nevertheless be
``acceptable''. For a process which does not
require high accuracy (e.g. the production of tissue)
the acceptable bounds will be fairly wide. In contrast,
for processes which require high accuracy they may be rather narrow.
\qpar

Although we know of few specific examples of defect forming
here is one. \qL
There is a mechanism called the ``induced-fit
process'' but that we prefer to call the ``hand-glove adjustment''
in which a substrate must do more than simply fit
into the already preformed shape of an enzyme's active site.
When the substrate approaches the enzyme surface, it
induces a change in its shape that optimizes a
correct placement. In short, there is a two-sided adjustment,
hence the hand-glove expression.
A case in point is the digestive enzyme carboxypeptidase in which
the binding of the substrate causes a tyrosine
molecule (a standard hydrophobic
amino acid used by cells to synthesize
proteins)
at the active site to move by as much as 15 angstroms
which is about twice the size of the molecule itself. In fact,
depending on the temperature, the
real distance will differ. A large change of that kind may even disrupt the
adjustment and therefore the production of the protein.
\qpar

We see that although examples of mishaps can be given, we have
little intuitive understanding of them because these phenomena occur
in the microscopic world with which we are not familiar

\qA{Mutations}

In contrast, we have a fairly clear view of the effects of
DNA mutations because these are illustrated in daily experiments
in which one can describe side by side the mutation and its
phenotypic consequences%
\qfoot{Whereas the connection mutation $ \rightarrow $ abnormality
is easy, the converse connection is much more difficult. In fact
the genetic basis of most abnormalities is unknown. }%
.
In other words it is not surprising that 
we are much more familiar with mutations than
with mishaps.
By the way, the same holds in our building parallel. Whereas
it is very
easy to identify a flaw with respect to the blueprint,
a defect in the structural properties of the walls is not
easily detected.

\qA{Randomness versus semi-randomness: the case of strabism}

Finally, one must keep in mind one major difference
between mishaps and mutations: while mutations are
completely random, mishaps are rather semi-random.
In which sense do we mean that? Depending upon the degree of
accuracy that is required, mishaps will be more or less
frequent. Here is an illustration.
\qpar

The control
of the pupils of the eyes requires an excellent coordination of
the surrounding muscles; this means that only small fluctuations
are acceptable, which in turn imposes
narrow bounds on the values of external
parameters. Thus, one will not
be surprised that this congenital malformation (called strabism)
occurs with a
fairly high frequency of the order of 3\% according to US statistics.
This is one of the highest malformation frequencies.
Similarly, the making of heart valves requires strict manufacturing
constraints; as a result heart valve defects have a relatively high
likelihood.

\qI{Appendix B. Comparison of experimental protocols}

In reviewing the literature we found that the experimental conditions
are often loosely chosen and described. As a particularly striking
case one can mention the following (IAEA 1973, p. 55, Fig.10).
The graph shows two curves (a) and (b)
of surviving bacteria fractions as functions of the 
radiation dose.  The caption says: ``S. typhimurium irradiated in meat
at -15C. 
(a) Following growth in the meat at 37C for two days. 
(b) Without pre-irradiation growth in the meat''. 
The growth rates before irradiatation
are not given. 
In (a) the culture was brutally brought from a temperature of 37C to
-15C. For (b) the pre-irradiation temperature is not indicated.
In short, this graph mixes up 
temperature  and radiation effects. Due to an unclear design and a
sketchy description it is impossible to draw any conclusion. 
\qpar

This example shows that great care should be given to 
both the design and the description of the experimental conditions.
It is the purpose of this appendix to emphasize some important
distinctions in this respect.

\qA{Three types of measurements}

Before we come to the data we wish to describe how
these experiments can be done. There are basically three methods.
\qee{{\bf Type 1}} The simplest way is to use bacteria which have
different {\it generation times} $ G $. This is the time 
interval
between successive divisions as long as one is in the
exponential phase. In this phase, the population 
doubles in time $ G $ which for that reason is also
called the doubling time. As examples, 
for the bacteria {\it Salmonella typhimurium} $ G $
is equal to 20mn, whereas for the bacteria 
{\it Micrococcus radiodurans} $ G $ is equal to
5 hours (i.e. 15 times more).
\qee{{\bf Type 2}} If one wishes to focus on a specific bacterium
one can make a comparison between $ G $ in the exponential
phase and the very large value of $ G $ observed in
the stationary phase. Although many papers use
the comparison exponential phase  versus stationary phase
one should observe that it is a methodology which
is not very accurate for at least two reasons. \qL
(i) The comparison relies on only two values. It is true
that in addition $ G $ can be changed by controlling the temperature, 
however we did not find any paper in which
this possibility is used. \qL
(ii) Actually $ G $ is not really constant in the 
exponential phase; it increases slowly but steadily.
In addition, for bacteria like {\it E. coli} which have a small
$ G $ the whole exponential phase lasts less than 3 hours
which may be too short to fully observe the impact
of the exogenous factor.
\qee{{\bf Type 3}} The best method is to use chemostat-cultivated 
bacteria. A chemostat is a device in which a flow is
maintained so that the bacteria remain in the exponential
phase indefinitely. In addition, it is possible to control
the growth rate by changing the dilution rate. Unfortunately,
the chemostat methodology requires an appropriate device
and an initiation to learn how to control the various parameters.
In the paper by Linqvist et al. (2014) there is a useful  comparison,
for the same stress factor,
between a batch  and a chemostat experiment.

\qA{Two key-variables and how to measure them}

These key-variables can be described as follows:
\qee{(a)} The overall number of cells, whether dead or alive, 
is measured with a spectrophotometer which tests the turbidity of the
medium. Increases in the optical density give estimates of the
number of cell divisions, and therefore of the growth rate in the
form $ \alpha=(1/\Delta t)\log[y(t)/y_0] $ 
which is constant for an exponential
growth. Note that the doubling time is given by: 
$ \tau=\log2/\alpha $.
\qee{(b)} The second key-variable is the number of viable cells.
It is estimated either with a cytometer or through the CFU
(colony forming unit) method. When a cell can give rise to a
colony on a Petri dish it is considered viable. Note that some 
living cells may become unable to develop into colonies.
For this and other reasons the method is fairly
inaccurate. Wilson (1921) gives many examples of possible
pitfalls.

\vskip 5mm

{\bf Acknowledgements} We are grateful to 
Nicolas Biais (Sorbonne University) for
a discussion which was at the origin of this paper. One of us (BR)
had the privilege of many stimulating exchanges with Arnaud Chastanet (INRAE)
and he wishes to extend many thanks to him.

\vskip 10mm

{\bf References}

\qparr
Benito (A.), Ventoura (G.), Casadei (M.), Robinson (T.),
Mackey (B.) 1999: Variation in resistance of natural
isolates of {\it Escherichia coli} O157 to high hydrostatic
pressure, mild heat and other stresses.
Applied and Environmental Microbiology 65,1564-1569.

\qparr
Berrut (S.), Pouillard (V.), Richmond (P.), 
Roehner (B.M.) 2016: Deciphering infant mortality.
Physica A 463,400-426.

\qparr
Berney(M.), Weilenmann (H.-U.), Ihssen (J.), Bassin (C.),
Egli (T.) 2006: Specific growth rate determines the
sensitivity of {\it Escherichia coli}  to thermal, UVA, and
solar disinfection.
Applied and Environmental Microbiology April, p. 2586-2593.

\qparr
Chen (Q.), Di (Z.), Garcia-Roger (E.M.),
Li (H.), Richmond (P.), Roehner (B.M.) 2020:
Magnitude and significance of the peak of early
embryonic mortality.
Journal of Biological Physics, 17 August 2020.

\qparr
IAEA (International Atomic Energy Agency) 1973: Manual
on radiation sterilization of medical and biological
materials. Vienna.

\qparr
Ihssen (J.), Egli (T.) 2004: Specific growth and not
cell density controls the general stress response in
{\it Escherichia coli}. 
Microbiology 150, 1637-1648.

\qparr
Lacassagne (M.A.) 1930: Diff\'erence de l'action biologique
provoqu\'ees  dans les levures par diverses radiations. 
[Action of X rays, $ \alpha $ rays and ultraviolet light on yeast]
Comptes Rendus de l'Acad\'emie des Sciences, Paris, 190,524--526.

\qparr
Lea (D.E.) 1947: Actions of radiations on living cells.
Macmillan, New York. 

\qparr
Lindqvist (R.), Barmark (G.) 2014: 
Specific growth rate determines the
sensitivity of {\it Eschrichia coli} to lactic acid stress.
Implications for predictive microbiology.
BioMed Research International. Article 471317.

\qparr
Lwoff (A.) 1959: Factors influencing the evolution of viral
diseases at the cellular level and in the organism.
Bacteriological Reviews 23,3,109--124.

\qparr
Manas (P.), Mackey (B.M.) 2004: Morphological
and physiogical changes induced by high hydrostatic
pressure in exponential and stationary phase cells of
{\it Eschrichia coli}: relationship with cell death.
Applied and Environmental Microbiology March 1545-1554.

\qparr
Richmond (P.), Roehner (B.M.) 2016a:
Effect of marital status on death rates. Part1: High accuracy
exploration of the Farr--Bertillon effect.
Physica A 450,748-767.

\qparr
Richmond (P.), Roehner (B.M.) 2016b: 
Effect of marital status on death rates. Part 2: Transient
mortality spikes.
Physica A 450,768-784.

\qparr
Serianni (R.W.), Bruce (A.K.) 1968: Radioresistance
of {\it Micrococcus radiodurans} during the growth cycle.
Radiation Research 36,193-207.

\qparr
Shehata (T.E.), Marr (A.G.) 1971: Effect of nutrient concentration on 
the growth of {\it Escherichia coli.}\qL
[This article shows that the response of cells to the
concentration of
essential nutrients (e.g. glucose,
phosphate or tryptophan) is of the 0-1 type, meaning that
under a critical threshold value the cells cannot develop
whereas they grow normally for concentrations above that value.]

\qparr
Sherman (J.M.), Albus (W.R.) 1923: Physiological youth in bacteria.
Journal of Bacteriology 8,127-139.

\qparr
Soufi (B.), Krug (K.), Harst (A.), Macek (B.) 2015: 
Characterisation of the {\it E. coli} proteome and its modification
during growth and ethanol stress.
Frontiers in Microbiology 18 February.

\qparr
Wilson (G.S.) 1921: The proportion of viable bacteria in
young cultures with especial reference to the technique 
employed in counting.
Journal of Bacteriology 7,4,405-446.

\end{document}